\documentclass[aps,prl,reprint,10pt]{revtex4-1}
\usepackage{graphicx} 
\usepackage{comment}
\pdfoutput=1
\begin{document}
\title{Leptonic CP problem in left-right symmetric model}
\author{Ravi Kuchimanchi}
\email{ravi@aidindia.org}
\begin{abstract}
We find using the minimal left-right symmetric model that the presence of leptonic $CP$ violation can radiatively generate a strong $CP$ phase at the one-loop level itself, which can be beyond the current bounds established by the neutron electic dipole moment experiments. If there are no axions or unnatural cancellations, this leads to the testable prediction that leptonic $CP$ violation must be negligibly small (Dirac phase $\delta_{CP} = 0$ or $\pi$), in wide and interesting regions of parameter space.   
\end{abstract}

\maketitle


\noindent \emph{Introduction --} 

One of the most attractive extensions of the standard model is the left-right symmetric model~\cite{PhysRevD.10.275,*PhysRevD.11.566,*Senjanovic:1975rk} based on $SU(3)_c \times SU(2)_L \times SU(2)_R \times U(1)_{B-L}\times P$ that  restores  parity as a good symmetry of the Lagrangian. The model requires the introduction of 3 right handed neutrinos, which are parity partners of the left handed neutrinos, and thereby provides a strong reason for neutrino masses and mixings. Interestingly, the QCD vacuum angle $\theta_{QCD}$ is absent in the left-right $P$ symmetric (LR) model as it is parity odd, and the strong $CP$ phase $\bar{\theta}$ (that contributes to the neutron's electric dipole moment) is calculable in terms of the other parameters of the Lagrangian. 

In the standard model, if we set the tree-level strong $CP$ parameter $\bar{\theta}$ to zero by hand, it is not produced radiatively till the third loop and is negligibly small ($\bar{\theta} \sim 10^{-16}$)~\cite{Ellis:1978hq}.   In the  left-right symmetric model, the parity-odd $\bar{\theta}$ would be zero had $P$ remained unbroken. However a single $CP$ violating quartic term in the Higgs potential 
can generate a large $\bar{\theta}$  at the tree level once parity is spontaneously broken. If there are no unnatural cancellations between tree-level and radiative contributions to $\bar{\theta}$, the coupling $\alpha_2$ of this quartic term must be nearly real (or $CP$ conserving), so that its tree-level contribution to $\bar{\theta}$ is within the experimental bound $\lesssim 10^{-10}$ (or zero). We can ask in which loop order it is generated from other phases.  
Just like in the standard model, it was shown that even in the left-right model, $\bar{\theta}$ is not generated up to the third loop~\footnote{Though~\cite{Gronau:1985sp} does not specifically mention the strong $CP$ phase, the radiatively generated $CP$ violating quartic term of the Higgs potential, obtained in the 4th loop order in that work, contributes to the Higgs bidoublet VEVs and generates a strong $CP$ phase.}. However this calculation (see last two paragraphs of Ref~\cite{Gronau:1985sp}) in the left-right model had only looked at $CP$ violating radiative corrections from the CKM phase of the quark sector. 

In this letter we show that $CP$ violation in the leptonic sector can radiatively generate a complex phase in $\alpha_2$ and thereby the strong $CP$ phase $\bar{\theta}$, at the one-loop level itself.  Moreover if the Dirac-type Yukawa terms in the leptonic sector are similar to their quark sector counterparts, for a wide region in parameter space that includes all of type 2 dominance, and some interesting regions of type 1 seesaw mechanism with right handed breaking scale $v_R \lesssim 10^{15} GeV$, the strong $CP$ phase generated from the leptonic phases exceeds the bound $\bar{\theta} \leq 10^{-10}$ set by neutron EDM experiments.  Thus we predict that leptonic $CP$ violation must be absent or unobservably small 
in the left-right symmetric model with the above provisions, and thereby we may be able to test if parity is restored in laws of nature at some scales well beyond collider reach. Moreover, if all the neutrino Dirac Yukawas are similar to the electron's Yukawa coupling, we find using Type 2 seesaw that $\delta_{CP} \lesssim 1/30$  if  $v_R \sim 1 TeV$, while for $v_R \sim 10 TeV$ an observable neutron EDM 
is generated. 

It is worth noting that there are axionless solutions to the strong $CP$ problem that restore $CP$ at a relaxation scale above $v_R$,  such that even after spontaneous or soft $CP$ breaking, $\alpha_2$ is naturally real at the relaxation or cut-off scale, while the $CKM$ phase is generated~\cite{Kuchimanchi:2010xs}. These solutions are discussed towards the end of the letter. 

That a complex $\alpha_2$ is generated from phases in the leptonic sector, should have shown up in the one-loop Renormalization Group Equations of the left-right symmetric model which were evaluated in~\cite{Rothstein:1990qx}. However the RG equations obtained in that work do not contain the contribution to the imaginary part of $\alpha_2$ (denoted by $\lambda_{11}$ in~\cite{Rothstein:1990qx} and $\alpha_{2I}$ in this work) from the phases in leptonic Yukawa matrices.  More recent work such as~\cite{Maiezza:2014ala} also does not find $CP$ violating one loop contributions to $\bar{\theta}$, if $\alpha_2$ is real or $CP$ conserving at tree level.   Our result is a significant departure from all previous works   which concluded or assumed that  $\alpha_{2I}$ (and therefore $\bar{\theta}$), once set to zero at the tree-level, does not pick up $CP$ violating radiative corrections at the one loop level in the LR model.  

The excessive one loop corrections imply that for the smallness of $\bar{\theta}$ to be natural in a technical sense, in important regions of parameter space of the non-supersymmetric LR model, the strong $CP$ problem \emph{must} be solved by introducing an axion, or else the leptonic $CP$ violating phases must be suppressed, which is testable.     


In supersymmetic models, that phases from \emph{soft} trilinear terms involving the sleptons can contribute to $\bar{\theta}$ in one loop was found in~\cite{Dedes:1999uj}.  However that work had to abandon  the calculation of $\bar{\theta}$ from leptonic phases in non-supersymmetric models with two Higgs doublets, since there are contributions to $\bar{\theta}$ from  imaginary parts of parameters of quadratically divergent dimension 2 Higgs mass terms, which \emph{have} to be fine tuned due to the Hierarchy problem (see~\cite{Dedes:1999uj}). However in the LR model all dimension 2 Higgs mass parameters are real due to $P$, and there is no imaginary part to quadratic divergences. Thus there is no roadblock to calculating the radiative contribution from leptonic phases, and it is well known that $\bar{\theta}$ is finite and calculable in the LR model with $P$. 


\vskip0.1in

\noindent \emph{Connection between Strong and Leptonic $CP$ violation --}

We consider the minimal Left-Right symmetric model~\cite{PhysRevD.10.275,PhysRevD.11.566,Senjanovic:1975rk} based on $G_{LR} \equiv SU(3)_c \times SU(2)_L \times SU(2)_R \times U(1)_{B-L} \times P$, with scalar triplets $\Delta_R$ (1, 1, 3, 2) and $\Delta_L$ (1, 3, 1, 2), and bi-doublet $\phi$ (1, 2, 2, 0). Under parity (P),  the space-time coordinates $(x,t) \rightarrow (-x,t), \ \phi \rightarrow \phi^\dagger$ and subscripts $L \leftrightarrow R$ for all other fields (see for example~\cite{Duka:1999uc}).   The scalar fields have the form
\begin{equation}
\begin{array}{ccc}
\phi = \left(\begin{array}{cc}
\phi^o_1 & \phi^+_2 \\
\phi^-_1 & \phi^o_2
\end{array}
\right),  &
\Delta_{L,R} = \left(\begin{array}{cc}
\delta^+_{L,R} / \sqrt{2} & \delta^{++}_{L,R} \\
\delta^o_{L,R} & - \delta^+_{L,R} /\sqrt{2}
\end{array}
\right).
\end{array}
\label{eq:fields}
\end{equation}

As is well known, all parameters of the Higgs potential are real due to $P$, except $\alpha_2$ in the $CP$ violating term~\cite{Duka:1999uc} 
\begin{eqnarray}
V= \left[\alpha_2 Tr\left(\Delta^\dagger_R \Delta_R\right) + \alpha^\star_2 Tr\left(\Delta^\dagger_L \Delta_L\right) \right] Tr\left(\tilde{\phi}^\dagger \phi\right) + h.c.
\label{eq:cpv}
\end{eqnarray}
where $\tilde{\phi} = \tau_2 \phi^\star \tau_2$.  If $\alpha_2$ is complex, once  $\delta^o_R$ picks up a vacuum expectation value (VEV) $\left<\delta^o_R\right>= v_R/\sqrt{2} >> \left<\delta^o_L\right>$, the above term generates a $CP$ violating coupling between the two standard model doublets in the bi-doublet $\phi$.  This causes the VEV $\left<\phi^o_2\right> \equiv k_2 e^{i\alpha}/\sqrt{2}$ of the neutral component of the second standard model doublet to pick up a phase ($\alpha$), where we have chosen a basis so that $\left<\phi^o_1\right> \equiv k_1/\sqrt{2}$ and $\left<\delta^o_R\right>$ are real, and the weak scale $v^2_{wk} = |k_1|^2 + |k_2|^2$. The up and down quark mass matrices ($M_u$ and $M_d$) are no longer Hermitian, as they are obtained from Yukawa couplings of the quarks that are Hermitian (due to P), and Higgs bi-doublet VEVs that are no longer all real.  The strong $CP$ phase 
\begin{equation}
\bar{\theta} = \arg \det(M_u M_d) \sim  (\alpha_{2I}/\alpha_3) (m_t/m_b)
\label{eq:theta}
\end{equation} 
therefore gets generated from the non-Hermitian mass matrices, and has been written in terms of the imaginary part of $\alpha_2 (= \alpha_{2R} + i \alpha_{2I})$, and the top to bottom quark mass ratio.  $\alpha_3$ is the real coupling of the term $\alpha_3 Tr(\phi^\dagger \phi\Delta_R \Delta^\dagger_R)$ that keeps the potential stable, and as noted in~\cite{Duka:1999uc} generates the mass $\sqrt{\alpha_3/2} v_R$ for the second standard model doublet.  Since experimentally~\cite{PhysRevLett.97.131801} $\bar{\theta} \lesssim 10^{-10}$, it is crucial that $\alpha_{2I}$ is close to zero to a high degree.  However we will show that, if $\alpha_2$ is chosen to be real at the tree level (with a cut-off scale $> v_R$),  a dangerous contribution to $\alpha_{2I}$ is generated at the one loop level  from the leptonic Yukawa terms given below:
\begin{eqnarray}
h^\ell_{ij} \bar{L}_{iL} \phi L_{jR} + \tilde{h}^\ell_{ij} \bar{L}_{iL} \tilde{\phi} L_{jR} + \nonumber \\ i f_{ij} \left(L^T_{iR}\tau_2 C \Delta_R L_{jR} + L^T_{iL}\tau_2 C \Delta_L L_{jL}\right) + H.c. 
\label{eq:yuka}
\end{eqnarray} 
where for example,  $\bar{L}_{1R}$ is the first generation leptonic doublet  $(\bar{\nu} \  \ \bar{e})_R$, and $3 \times 3$ Yukawa matrices $h^\ell$ and $\tilde{h}^\ell$ are Hermitian due to P, while $f$ is a complex, symmetric $3 \times 3$ matrix that generates Majorana terms for neutrinos.

\begin{figure}[t]
\begin{center}
\includegraphics[height=3.5cm]{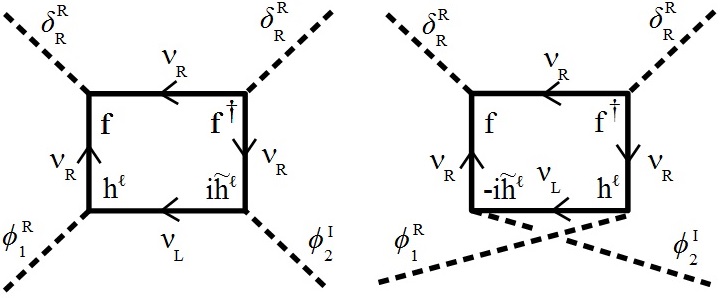}
\end{center}
\caption{One loop contribution to $\alpha_{2I}$ from leptonic Yukawas of eq.~(\ref{eq:yuka}), with $\phi^o_1 = \phi^R_{1} + i \phi^I_{1}, \ \phi^o_2 = \phi^R_{2} + i \phi^I_{2} $ and $\delta^{o}_R = \delta^R_R + i \delta^I_R$ in eq.~(\ref{eq:fields}). This leads to a VEV for $\phi^I_2$ and generates $\bar{\theta}$.}
\label{fig:alphaoneloop}
\end{figure}

The above Lagrangian radiatively generates a logarithmically divergent contribution to $\alpha_{2I}$ through the box diagrams of Figure~\ref{fig:alphaoneloop}, so that 
\begin{equation}
  \alpha_{2I} \sim \frac{i}{16 \pi^2} Tr\left( f^\dagger f \left[h^\ell, \tilde{h}^\ell\right]\right) \ln\left(v_R/M_{Pl}\right)^2 
\label{eq:oneloop}
\end{equation}
where we note that $i[h^\ell, \tilde{h}^\ell]$ and $f^\dagger f$ are Hermitian matrices, and the cut-off has been taken to be at the Planck scale $M_{Pl} > v_R$. 
Note that there  is no suppression in eq.~(\ref{eq:oneloop}) by a factor such as $(v_{wk}/v_R)^2$ or $(v_R/M_{Pl})^2$, and hence this contribution can be dangerously large, even if $v_R$ scale is well above the $TeV$ scale. 

Equation~(\ref{eq:oneloop}) with~(\ref{eq:theta}) generates $\bar{\theta}$ in the quark sector from leptonic Yukawas and $CP$ violation therein, and provides the following severe constraint on the leptonic sector that is being  probed by neutrino experiments:  
 \begin{equation}
\left|Tr\left(f^\dagger f \left[h^\ell, \tilde{h}^\ell\right]\right)\right| \lesssim 3 \times 10^{-11}
\label{eq:constraint}
\end{equation}
where we have substituted $\bar{\theta} \leq 10^{-10}, m_t/m_b \sim 40, \alpha_3 \lesssim 1$ and took the logarithm to have a generic value $\sim 10$. Taking the Hermitian conjugate of~(\ref{eq:constraint}) it can be seen that if $f^\dagger f, h^\ell$, and $\tilde{h}^\ell$ have all real matrix elements (conserve CP), then the left hand side vanishes. On the other hand, as we will now see,  if they have complex phases these can be constrained by the above equation.   

We will now consider the well motivated case where the leptonic Dirac Yukawa couplings are similar to their quark counterparts.  This for example would be the case if there is an ultra-violet completion with a semi-unified theory such as the Pati-Salam model with $SU(4)_c \times SU(2)_L \times SU(2)_R$, or grand unified theory such as $SO(10)$.  This implies that the matrix $[h^\ell, \tilde{h}^\ell]$ has some off-diagonal matrix elements that are of the order $\sim V_{ts} (m_b/m_t) \gtrsim 3 \times 10^{-4}$, in a basis in which either $h^\ell$ or $\tilde{h}^\ell$, like its quark counterpart is diagonal.  Even if there is no unification, that the quark and leptonic Dirac type Yukawa matrices could be similar is hinted by the fact that the charged lepton masses are similar to the down sector quark masses.  

With the above, equation~(\ref{eq:constraint}) implies that some off diagonal matrix elements 
\begin{equation}
\left|\left(f^\dagger f\right)_{ij}\right| \lesssim 10^{-7} 
\label{eq:ff}
\end{equation}
if there are $O(1)$ $CP$ phases present in $f^\dagger f$ or in $[h^\ell, \tilde{h}^\ell]$.

The Yukawa matrix $f$ leads to Majorana mass terms  for neutrinos once $\delta^o_R$ picks up a large VEV $\sim v_R/\sqrt{2}$ and $\delta^o_L$ picks up an induced VEV $\sim \gamma v^2_{wk} /v_R$ where $\gamma$ which is symbolically $\beta/\rho$ is obtained from real quartic couplings $\beta_i$ and $\rho_i$ of the Higgs potential~\cite{Duka:1999uc} (which has terms such as $\rho_1^2 Tr(\Delta^\dagger_R \Delta_R)^2 + R\rightarrow L$), and is real at tree level. Since no symmetries can protect $\rho_i$, we have in general, $\rho_i \gtrsim 0.01$ and therefore, $|\gamma| \lesssim 100$. The light neutrino mass matrix is given by the well known seesaw mechanism~\cite{Minkowski:1977sc,*van1979supergravity,*Yanagida01091980,*Mohapatra:1979ia}  and has the form
\begin{equation}
M_\nu = \frac{v^2_{wk}}{v_R} \left[\gamma f - h_D \left(\frac{1}{f}\right) h^T_D \right] 
\label{eq:seesaw}
\end{equation} 
where $h_D  = (k_1 h^\ell + k_2 e^{-i \alpha} \tilde{h}^\ell)/v_{wk}$ is the Dirac type Yukawa matrix for the neutrinos. 

If the first term in the square brackets of eq.~(\ref{eq:seesaw}) dominates over the second term, we have a Type 2 seesaw mechanism. Taking the third generation Yukawas to be larger than the rest of the Yukawas, for the first term to dominate, we must have $f_{33} > h_{D_{33}} / \sqrt{|\gamma|}$. Since we have assumed that leptonic Dirac type and quark Yukawas are similar, we take $h_{D_{33}} \sim h_t \gtrsim 0.3$.  Substituting $|\gamma| \lesssim 100$ we have for Type 2 seesaw mechanism, $f_{33} \gtrsim 0.03$.  Since $M_{\nu} \approx f\gamma v^2_{wk}/v_R$ for type 2 seesaw, and we know by light neutrino experiments that the leptonic mixing angles are large we have for the off diagonal matrix elements, $f_{3j} \sim f_{33} \ to \ f_{33}/10$. Thus we obtain $|(f^\dagger f)_{3j}| \gtrsim 10^{-4}$.

Comparing the above with eq.~(\ref{eq:ff}) we can see that the leptonic Dirac phase $\delta_{CP}$ cannot be of the order $1$, and  must be less than $10^{-3}$ from its $CP$ conserving value of $0$ or $\pi$.  Note that in this case of type 2 seesaw, the Majorana $CP$ violating phases are unconstrained since they do not occur in $M^\dagger_\nu M_\nu \propto f^\dagger f$ or in $[h^\ell,\tilde{h}^\ell]$.  

Currently experiments are being planned or underway to measure $\delta_{CP}$ with a sensitivity of $5^o$ (or $\sim 0.1$)~\cite{deGouvea:2013onf}, similar to sensitivity achieved for the $CKM$ phase. The absence of a measurable $\delta_{CP}$ (modulo $\pi$) for the above well motivated case, is a key prediction of this work.  

We now consider the case of type 1 seesaw where the second term in eq.~(\ref{eq:seesaw}) dominates.  Substituting $h_{D_{33}} \sim 0.3$ we find that for type 1 seesaw, $f_{33} v_R \sim 10^{14} GeV$ so that $h^2_{D_{33}} v^2_{wk} /(f_{33}v_R) \sim \sqrt{|\Delta m^2_{32}|} \sim 0.05 eV$, where we have used the mass squared difference of light neutrinos~\cite{deGouvea:2013onf} $|\Delta m^2_{32}| \sim 0.0023eV^2$ and $v_{wk} \sim 246 GeV$.  

If $v_R \sim 10^{18} GeV$, we must have $f_{33} \sim 10^{-4}$ and we can see that the off-diagonal terms of $f^\dagger f$ will satisfy eq.~(\ref{eq:ff}), and there is no constraint on the leptonic $CP$ phases. On the other hand if $v_R \sim 10^{15 \ to \ 14}GeV$, then $f_{33} \sim 0.1 \ to \ 1$.  For type 1 seesaw since $f$ is more hierarchical, we take $f_{23} \sim f_{33}/1000$ ($f_{23}$ would be about $f_{33}/100$ but we allow for an additional factor of 10 since the phases in $f_{23}$ could be order $\delta_{CP}/10$).  Comparing now with eq.~(\ref{eq:ff}) we once again find that leptonic $CP$ phases cannot be order 1 and must be constrained to be $\lesssim 10^{-(2 \ to \ 4)}$, modulo $\pi$. This is another key prediction.  Note also that for type 1 seesaw, since both Majorana and Dirac phases can be present in $f^\dagger f$, all the $CP$ phases are constrained.

So far we have looked at cases where quark and leptonic Dirac Yukawas are similar.  We now relax this assumption to consider an example where $v_R$ can be at the $TeV$ scale.  In a basis where charged lepton masses are diagonal, if all neutrino Dirac Yukawas including those of the third generation such as $h_{D_{3j}} \sim 10^{-5}$ to $10^{-6} \sim h^\ell_{3j}$, so that they are similar to the smallest known (electron's) Yukawa coupling, then $v_R$ can be $\sim 1 TeV$. This determines $f_{33} \sim 1$, since the second term in eq.~(\ref{eq:seesaw}) should not give a contribution much greater than the observed $0.0023 eV^2$for light neutrino mass-squared differences. Moreover, $[h^\ell, \tilde{h}^\ell]$ can have off-diagonal elements of the order $10^{-6} \times 10^{-2} = 10^{-8}$, since both $h_D$ and  charged fermion masses (with Yukawa of $\tau^-,  h_\tau \sim 10^{-2} \sim \tilde{h}^\ell_{33} $) must arise from  $h^\ell$ and $\tilde{h}^\ell$. For type 2 seesaw,  $f_{3j}\sim f_{33}/10 \sim 1/10$ and so we find that the left hand side of eq.~(\ref{eq:constraint}) is $\sim 10^{-9} \delta_{CP}$ and therefore $\delta_{CP}$ cannot be order 1 and is $\sim 1/30$.  However if $v_R \sim 10 TeV$ (so that $f_{33} \gtrsim 0.1$),  then $\delta_{CP} \sim 1$ maybe allowed and can result in an observable neutron EDM ($\bar{\theta} \gtrsim 3 \times 10^{-11} \delta_{CP}$). Null results in future nEDM experiments probing $\bar{\theta} \sim 10^{-12}$ may further constrain $\delta_{CP}$. 
\vskip0.1in
\noindent \emph{Strong $CP$ solution --}
We have  not assumed anything beyond the minimal left-right symmetric model to obtain the connection between  leptonic and strong $CP$ violation. 
It is clear that if the leptonic $CP$ phases are $O(1)$, then $\bar{\theta}$ may have to be fine tuned to cancel excessive one-loop radiative corrections, making it technically unnatural. On the other hand if leptonic phases and $\alpha_{2I}$ are zero at the tree level, there could be an underlying symmetry reason.   This  motivates us to look at solutions to the strong $CP$ problem.

If we extend the model by adding an axion, then $\bar{\theta}$ dynamically relaxes to zero~\cite{PhysRevLett.38.1440}.  However since $P$  sets the QCD vacuum angle to zero, historically it has been hoped that there would be an axionless solution in the LR model. An early attempt was made in~\cite{Mohapatra:1978fy} by adding an additional bi-doublet, and invoking a discrete symmetry. However after symmetry breaking the $CKM$ phase is not generated, and the problem remained unsolved. Later it was noted that  $\alpha_2$ is automatically absent in SUSYLR models and the strong $CP$ problem can be thus solved~\cite{Kuchimanchi:1995rp,*Mohapatra:1995xd}.  However it was shown for the SUSYLR solution, that without any further constraints, the radiatively generated $\bar{\theta} \sim 10^{-8} \ to \ 10^{-10}$~\cite{Babu:2001se} which is uncomfortably close to the experimental bound, while a solution in the LR model (without supersymmetry), continued to be elusive~\cite{Mohapatra:1997su} at the turn of the century.     

Recently, progress was made by adding one heavy vectorlike quark family to the minimal LR model, and breaking both $P$ and $CP$ spontaneously so that $\alpha_{2I}$ naturally vanishes at the $CP$ restoration scale~\cite{Kuchimanchi:2010xs}. 
$CP$ is spontaneously broken by the VEV of a $P$ even, $CP$ odd real scalar singlet whose Yukawa couplings mix the  usual and vectorlike quarks and generate the CKM phase. 
Since  $P$ is not broken by the singlet VEV, the resultant tree-level quark mass matrices are Hermitian, thus solving the strong $CP$ problem without requiring supersymmetry. 
The solution also works without the scalar singlet, if $CP$ is broken softly by mass terms involving the vectorlike quarks~\cite{Kuchimanchi:2010xs}.

The interesting thing is that in the minimal version of the above solution, since a vectorlike lepton family is not introduced, no $CP$ violation is generated in the lepton sector!  Thus it predicts that not only $\alpha_{2I}$ but also the Dirac ($\delta_{CP}$) and Majorana leptonic phases vanish (modulo $\pi$), as noted in ~\cite{Kuchimanchi:2012xb} and detailed in~\cite{Kuchimanchi:2012te}.  It is remarkable that the solution addresses perfectly the issues raised in this work.

However if a vectorlike lepton family is introduced then $\delta_{CP}$ can be generated.  This work shows that it would in turn generate too high a $\bar{\theta}$, in wide and interesting regions of parameter space, and hence the solution that includes a vectorlike lepton family may be disfavored.


\vskip0.1in
\noindent \emph{Neutron EDM in axionless LR solution -- } If $v_R << M$,  we just have the minimal LR model below $M$ (mass of the vector-like quark family or equivalently the scale of $CP$ breaking). Radiative corrections from heavy quarks introduce a slight non-Hermiticity in the light quark  mass matrices  and generate a finite and calculable $\bar{\theta}$.  
This was estimated in~\cite{Kuchimanchi:2010xs,Kuchimanchi:2012xb}, and it was found that contribution from terms at the  one-loop level are of the form,  
\begin{equation}
\bar{\theta} \sim  \frac{1}{16\pi^2}(Product \ of \ Yukawas) \left[\frac{v_R}{M}\right]^2
\end{equation}
where the term in the round brackets is essentially a product of a string of up and down quark Yukawa matrices with the standard model Higgs doublet, and includes one Yukawa inverse. The two loop contribution has terms of a similar form with a longer Yukawa string in the product and an additional factor of $(4\pi)^{-2}$. 

If $M \sim M_{Pl} \sim 10^{18}GeV$ and $v_R \sim 10^{14}$ to $10^{15}GeV$, then the above implies $\bar{\theta} \sim 10^{-12}$ to $10^{-10}$ if the  product of Yukawas in the round brackets is $\sim h^2_t V_{ts} \sim 1/100$. If some unknown Yukawas involving heavy quarks are smaller, it can further reduce $\bar{\theta}$.

An important feature in the above radiative corrections due to the heavy quarks is the suppression factor $(v_R/M)^2$. This is because as $M \rightarrow \infty$ the vector like quarks decouple.   
 Additionally, if there are Planck scale corrections due to non-renormalizable terms, they will induce a $\bar{\theta}_{Pl} \sim \lambda v^2_R/(M M_{Pl})$, where $\lambda$ is a dimensionless parameter and $M$ is the soft $CP$ breaking scale for the minimal Higgs content without a singlet. Thus the radiatively generated $\bar{\theta}$ due to vector like quarks, and  Planck scale corrections are both suppressed by $(v_R / M_{Pl})^2$ for $M \sim M_{Pl}$. Even if $v_R$ is small enough that both are undetectable, we still have the testable prediction that leptonic phases vanish (modulo $\pi$) in the axionless model.       

\vskip0.1in
\noindent \emph{Conclusions --} 
We have shown that presence of leptonic $CP$ violation can radiatively generate an excessive strong $CP$ phase at the one loop level in the minimal left-right symmetric model with $P$, that is beyond the  limits already established by the neutron EDM experiments for the following interesting regions of parameter space:
\begin{itemize}
\item if the leptonic and quark Dirac type Yukawa couplings are similar then 
\begin{itemize} 
\item for all regions of Type 2 dominant seesaw.  
\item for $v_R \lesssim 10^{15}GeV$ with Type 1 seesaw.
\end{itemize}
\item if the Dirac Yukawa couplings of the neutrinos, $h_{D_{3j}} \sim 10^{-5}$ to $10^{-6}$, so that $v_R \sim 1~TeV$, then for Type 2 seesaw.  
\end{itemize}  
In the above significant regions of parameter space, technical naturalness implies that leptonic $CP$ phases (particularly $ \delta_{CP}$)  must vanish or be negligibly small (modulo $\pi$), unless there are axions. The result is important as it gives us a way to test if parity is restored in the laws of nature, even if it is at scales $\sim 10^{15} GeV$ that are well beyond collider reach, in some well motivated regions of parameter space of the axionless minimal LR model.  In  the second bullet point of the above, if $v_R \sim 10~TeV$ then an observable neutron EDM ($\bar{\theta} \gtrsim 3 \times 10^{-11} \delta_{CP}$) is generated.  



In general, we can think of a theory as being afflicted with a leptonic $CP$ problem if leptonic phases generate an excessive strong $CP$ phase radiatively or through RGE running from higher scales. 

\begin{acknowledgments}
I thank Rabi Mohapatra and Kaustubh Agashe for helpful comments on the manuscript.
\end{acknowledgments}

\bibliography{LR1015_bibtex}

\begin{thebibliography}{25}%
\makeatletter
\providecommand \@ifxundefined [1]{%
 \@ifx{#1\undefined}
}%
\providecommand \@ifnum [1]{%
 \ifnum #1\expandafter \@firstoftwo
 \else \expandafter \@secondoftwo
 \fi
}%
\providecommand \@ifx [1]{%
 \ifx #1\expandafter \@firstoftwo
 \else \expandafter \@secondoftwo
 \fi
}%
\providecommand \natexlab [1]{#1}%
\providecommand \enquote  [1]{``#1''}%
\providecommand \bibnamefont  [1]{#1}%
\providecommand \bibfnamefont [1]{#1}%
\providecommand \citenamefont [1]{#1}%
\providecommand \href@noop [0]{\@secondoftwo}%
\providecommand \href [0]{\begingroup \@sanitize@url \@href}%
\providecommand \@href[1]{\@@startlink{#1}\@@href}%
\providecommand \@@href[1]{\endgroup#1\@@endlink}%
\providecommand \@sanitize@url [0]{\catcode `\\12\catcode `\$12\catcode
  `\&12\catcode `\#12\catcode `\^12\catcode `\_12\catcode `\%12\relax}%
\providecommand \@@startlink[1]{}%
\providecommand \@@endlink[0]{}%
\providecommand \url  [0]{\begingroup\@sanitize@url \@url }%
\providecommand \@url [1]{\endgroup\@href {#1}{\urlprefix }}%
\providecommand \urlprefix  [0]{URL }%
\providecommand \Eprint [0]{\href }%
\providecommand \doibase [0]{http://dx.doi.org/}%
\providecommand \selectlanguage [0]{\@gobble}%
\providecommand \bibinfo  [0]{\@secondoftwo}%
\providecommand \bibfield  [0]{\@secondoftwo}%
\providecommand \translation [1]{[#1]}%
\providecommand \BibitemOpen [0]{}%
\providecommand \bibitemStop [0]{}%
\providecommand \bibitemNoStop [0]{.\EOS\space}%
\providecommand \EOS [0]{\spacefactor3000\relax}%
\providecommand \BibitemShut  [1]{\csname bibitem#1\endcsname}%
\let\auto@bib@innerbib\@empty
\bibitem [{\citenamefont {Pati}\ and\ \citenamefont
  {Salam}(1974)}]{PhysRevD.10.275}%
  \BibitemOpen
  \bibfield  {author} {\bibinfo {author} {\bibfnamefont {J.~C.}\ \bibnamefont
  {Pati}}\ and\ \bibinfo {author} {\bibfnamefont {A.}~\bibnamefont {Salam}},\
  }\href {\doibase 10.1103/PhysRevD.10.275} {\bibfield  {journal} {\bibinfo
  {journal} {Phys. Rev. D}\ }\textbf {\bibinfo {volume} {10}},\ \bibinfo
  {pages} {275} (\bibinfo {year} {1974})}\BibitemShut {NoStop}%
\bibitem [{\citenamefont {Mohapatra}\ and\ \citenamefont
  {Pati}(1975)}]{PhysRevD.11.566}%
  \BibitemOpen
  \bibfield  {author} {\bibinfo {author} {\bibfnamefont {R.~N.}\ \bibnamefont
  {Mohapatra}}\ and\ \bibinfo {author} {\bibfnamefont {J.~C.}\ \bibnamefont
  {Pati}},\ }\href {\doibase 10.1103/PhysRevD.11.566} {\bibfield  {journal}
  {\bibinfo  {journal} {Phys. Rev. D}\ }\textbf {\bibinfo {volume} {11}},\
  \bibinfo {pages} {566} (\bibinfo {year} {1975})}\BibitemShut {NoStop}%
\bibitem [{\citenamefont {Senjanovic}\ and\ \citenamefont
  {Mohapatra}(1975)}]{Senjanovic:1975rk}%
  \BibitemOpen
  \bibfield  {author} {\bibinfo {author} {\bibfnamefont {G.}~\bibnamefont
  {Senjanovic}}\ and\ \bibinfo {author} {\bibfnamefont {R.~N.}\ \bibnamefont
  {Mohapatra}},\ }\href {\doibase 10.1103/PhysRevD.12.1502} {\bibfield
  {journal} {\bibinfo  {journal} {Phys. Rev.}\ }\textbf {\bibinfo {volume}
  {D12}},\ \bibinfo {pages} {1502} (\bibinfo {year} {1975})}\BibitemShut
  {NoStop}%
\bibitem [{\citenamefont {Ellis}\ and\ \citenamefont
  {Gaillard}(1979)}]{Ellis:1978hq}%
  \BibitemOpen
  \bibfield  {author} {\bibinfo {author} {\bibfnamefont {J.~R.}\ \bibnamefont
  {Ellis}}\ and\ \bibinfo {author} {\bibfnamefont {M.~K.}\ \bibnamefont
  {Gaillard}},\ }\href {\doibase 10.1016/0550-3213(79)90297-9} {\bibfield
  {journal} {\bibinfo  {journal} {Nucl.Phys.}\ }\textbf {\bibinfo {volume}
  {B150}},\ \bibinfo {pages} {141} (\bibinfo {year} {1979})}\BibitemShut
  {NoStop}%
\bibitem [{Note1()}]{Note1}%
  \BibitemOpen
  \bibinfo {note} {Though~\cite {Gronau:1985sp} does not specifically mention
  the strong $CP$ phase, the radiatively generated $CP$ violating quartic term
  of the Higgs potential, obtained in the 4th loop order in that work,
  contributes to the Higgs bidoublet VEVs and generates a strong $CP$
  phase.}\BibitemShut {Stop}%
\bibitem [{\citenamefont {Gronau}\ and\ \citenamefont
  {Mohapatra}(1986)}]{Gronau:1985sp}%
  \BibitemOpen
  \bibfield  {author} {\bibinfo {author} {\bibfnamefont {M.}~\bibnamefont
  {Gronau}}\ and\ \bibinfo {author} {\bibfnamefont {R.~N.}\ \bibnamefont
  {Mohapatra}},\ }\href {\doibase 10.1016/0370-2693(86)90973-1} {\bibfield
  {journal} {\bibinfo  {journal} {Phys.Lett.}\ }\textbf {\bibinfo {volume}
  {B168}},\ \bibinfo {pages} {248} (\bibinfo {year} {1986})}\BibitemShut
  {NoStop}%
\bibitem [{\citenamefont {Kuchimanchi}(2010)}]{Kuchimanchi:2010xs}%
  \BibitemOpen
  \bibfield  {author} {\bibinfo {author} {\bibfnamefont {R.}~\bibnamefont
  {Kuchimanchi}},\ }\href {\doibase 10.1103/PhysRevD.82.116008} {\bibfield
  {journal} {\bibinfo  {journal} {Phys. Rev.}\ }\textbf {\bibinfo {volume}
  {D82}},\ \bibinfo {pages} {116008} (\bibinfo {year} {2010})},\ \Eprint
  {http://arxiv.org/abs/1009.5961} {arXiv:1009.5961 [hep-ph]} \BibitemShut
  {NoStop}%
\bibitem [{\citenamefont {Rothstein}(1991)}]{Rothstein:1990qx}%
  \BibitemOpen
  \bibfield  {author} {\bibinfo {author} {\bibfnamefont {I.}~\bibnamefont
  {Rothstein}},\ }\href {\doibase 10.1016/0550-3213(91)90536-7} {\bibfield
  {journal} {\bibinfo  {journal} {Nucl.Phys.}\ }\textbf {\bibinfo {volume}
  {B358}},\ \bibinfo {pages} {181} (\bibinfo {year} {1991})}\BibitemShut
  {NoStop}%
\bibitem [{\citenamefont {Maiezza}\ and\ \citenamefont
  {Nemevšek}(2014)}]{Maiezza:2014ala}%
  \BibitemOpen
  \bibfield  {author} {\bibinfo {author} {\bibfnamefont {A.}~\bibnamefont
  {Maiezza}}\ and\ \bibinfo {author} {\bibfnamefont {M.}~\bibnamefont
  {Nemevšek}},\ }\href {\doibase 10.1103/PhysRevD.90.095002} {\bibfield
  {journal} {\bibinfo  {journal} {Phys.Rev.}\ }\textbf {\bibinfo {volume}
  {D90}},\ \bibinfo {pages} {095002} (\bibinfo {year} {2014})},\ \Eprint
  {http://arxiv.org/abs/1407.3678} {arXiv:1407.3678 [hep-ph]} \BibitemShut
  {NoStop}%
\bibitem [{\citenamefont {Dedes}\ and\ \citenamefont
  {Pospelov}(2000)}]{Dedes:1999uj}%
  \BibitemOpen
  \bibfield  {author} {\bibinfo {author} {\bibfnamefont {A.}~\bibnamefont
  {Dedes}}\ and\ \bibinfo {author} {\bibfnamefont {M.}~\bibnamefont
  {Pospelov}},\ }\href {\doibase 10.1103/PhysRevD.61.116010} {\bibfield
  {journal} {\bibinfo  {journal} {Phys.Rev.}\ }\textbf {\bibinfo {volume}
  {D61}},\ \bibinfo {pages} {116010} (\bibinfo {year} {2000})},\ \Eprint
  {http://arxiv.org/abs/hep-ph/9912293} {arXiv:hep-ph/9912293 [hep-ph]}
  \BibitemShut {NoStop}%
\bibitem [{\citenamefont {Duka}\ \emph {et~al.}(2000)\citenamefont {Duka},
  \citenamefont {Gluza},\ and\ \citenamefont {Zralek}}]{Duka:1999uc}%
  \BibitemOpen
  \bibfield  {author} {\bibinfo {author} {\bibfnamefont {P.}~\bibnamefont
  {Duka}}, \bibinfo {author} {\bibfnamefont {J.}~\bibnamefont {Gluza}}, \ and\
  \bibinfo {author} {\bibfnamefont {M.}~\bibnamefont {Zralek}},\ }\href
  {\doibase 10.1006/aphy.1999.5988} {\bibfield  {journal} {\bibinfo  {journal}
  {Annals Phys.}\ }\textbf {\bibinfo {volume} {280}},\ \bibinfo {pages} {336}
  (\bibinfo {year} {2000})},\ \Eprint {http://arxiv.org/abs/hep-ph/9910279}
  {arXiv:hep-ph/9910279} \BibitemShut {NoStop}%
\bibitem [{\citenamefont {Baker}\ \emph {et~al.}(2006)\citenamefont {Baker},
  \citenamefont {Doyle}, \citenamefont {Geltenbort}, \citenamefont {Green},
  \citenamefont {van~der Grinten}, \citenamefont {Harris}, \citenamefont
  {Iaydjiev}, \citenamefont {Ivanov}, \citenamefont {May}, \citenamefont
  {Pendlebury}, \citenamefont {Richardson}, \citenamefont {Shiers},\ and\
  \citenamefont {Smith}}]{PhysRevLett.97.131801}%
  \BibitemOpen
  \bibfield  {author} {\bibinfo {author} {\bibfnamefont {C.~A.}\ \bibnamefont
  {Baker}}, \bibinfo {author} {\bibfnamefont {D.~D.}\ \bibnamefont {Doyle}},
  \bibinfo {author} {\bibfnamefont {P.}~\bibnamefont {Geltenbort}}, \bibinfo
  {author} {\bibfnamefont {K.}~\bibnamefont {Green}}, \bibinfo {author}
  {\bibfnamefont {M.~G.~D.}\ \bibnamefont {van~der Grinten}}, \bibinfo {author}
  {\bibfnamefont {P.~G.}\ \bibnamefont {Harris}}, \bibinfo {author}
  {\bibfnamefont {P.}~\bibnamefont {Iaydjiev}}, \bibinfo {author}
  {\bibfnamefont {S.~N.}\ \bibnamefont {Ivanov}}, \bibinfo {author}
  {\bibfnamefont {D.~J.~R.}\ \bibnamefont {May}}, \bibinfo {author}
  {\bibfnamefont {J.~M.}\ \bibnamefont {Pendlebury}}, \bibinfo {author}
  {\bibfnamefont {J.~D.}\ \bibnamefont {Richardson}}, \bibinfo {author}
  {\bibfnamefont {D.}~\bibnamefont {Shiers}}, \ and\ \bibinfo {author}
  {\bibfnamefont {K.~F.}\ \bibnamefont {Smith}},\ }\href {\doibase
  10.1103/PhysRevLett.97.131801} {\bibfield  {journal} {\bibinfo  {journal}
  {Phys. Rev. Lett.}\ }\textbf {\bibinfo {volume} {97}},\ \bibinfo {pages}
  {131801} (\bibinfo {year} {2006})}\BibitemShut {NoStop}%
\bibitem [{\citenamefont {Minkowski}(1977)}]{Minkowski:1977sc}%
  \BibitemOpen
  \bibfield  {author} {\bibinfo {author} {\bibfnamefont {P.}~\bibnamefont
  {Minkowski}},\ }\href {\doibase 10.1016/0370-2693(77)90435-X} {\bibfield
  {journal} {\bibinfo  {journal} {Phys.Lett.}\ }\textbf {\bibinfo {volume}
  {B67}},\ \bibinfo {pages} {421} (\bibinfo {year} {1977})}\BibitemShut
  {NoStop}%
\bibitem [{\citenamefont {Gell-Mann}\ \emph {et~al.}(1979)\citenamefont
  {Gell-Mann}, \citenamefont {Ramond},\ and\ \citenamefont
  {Slansky}}]{van1979supergravity}%
  \BibitemOpen
  \bibfield  {author} {\bibinfo {author} {\bibfnamefont {M.}~\bibnamefont
  {Gell-Mann}}, \bibinfo {author} {\bibfnamefont {P.}~\bibnamefont {Ramond}}, \
  and\ \bibinfo {author} {\bibfnamefont {R.}~\bibnamefont {Slansky}},\ }\href
  {http://books.google.co.in/books?id=Dg-CAAAAIAAJ} {\emph {\bibinfo {title}
  {Supergravity}}},\ ed. by D. Freedman and P. Van Nieuwenhuizen, North
  Holland, Amsterdam, 315-321\ (\bibinfo {year} {1979})\BibitemShut {NoStop}%
\bibitem [{\citenamefont {Yanagida}(1980)}]{Yanagida01091980}%
  \BibitemOpen
  \bibfield  {author} {\bibinfo {author} {\bibfnamefont {T.}~\bibnamefont
  {Yanagida}},\ }\href {\doibase 10.1143/PTP.64.1103} {\bibfield  {journal}
  {\bibinfo  {journal} {Progress of Theoretical Physics}\ }\textbf {\bibinfo
  {volume} {64}},\ \bibinfo {pages} {1103} (\bibinfo {year}
  {1980})}\BibitemShut {NoStop}%
\bibitem [{\citenamefont {Mohapatra}\ and\ \citenamefont
  {Senjanovic}(1980)}]{Mohapatra:1979ia}%
  \BibitemOpen
  \bibfield  {author} {\bibinfo {author} {\bibfnamefont {R.~N.}\ \bibnamefont
  {Mohapatra}}\ and\ \bibinfo {author} {\bibfnamefont {G.}~\bibnamefont
  {Senjanovic}},\ }\href {\doibase 10.1103/PhysRevLett.44.912} {\bibfield
  {journal} {\bibinfo  {journal} {Phys.Rev.Lett.}\ }\textbf {\bibinfo {volume}
  {44}},\ \bibinfo {pages} {912} (\bibinfo {year} {1980})}\BibitemShut
  {NoStop}%
\bibitem [{\citenamefont {de~Gouvea}\ \emph {et~al.}(2013)\citenamefont
  {de~Gouvea} \emph {et~al.}}]{deGouvea:2013onf}%
  \BibitemOpen
  \bibfield  {author} {\bibinfo {author} {\bibfnamefont {A.}~\bibnamefont
  {de~Gouvea}} \emph {et~al.} (\bibinfo {collaboration} {Intensity Frontier
  Neutrino Working Group}),\ }\href@noop {} {\  (\bibinfo {year} {2013})},\
  \Eprint {http://arxiv.org/abs/1310.4340} {arXiv:1310.4340 [hep-ex]}
  \BibitemShut {NoStop}%
\bibitem [{\citenamefont {Peccei}\ and\ \citenamefont
  {Quinn}(1977)}]{PhysRevLett.38.1440}%
  \BibitemOpen
  \bibfield  {author} {\bibinfo {author} {\bibfnamefont {R.~D.}\ \bibnamefont
  {Peccei}}\ and\ \bibinfo {author} {\bibfnamefont {H.~R.}\ \bibnamefont
  {Quinn}},\ }\href {\doibase 10.1103/PhysRevLett.38.1440} {\bibfield
  {journal} {\bibinfo  {journal} {Phys. Rev. Lett.}\ }\textbf {\bibinfo
  {volume} {38}},\ \bibinfo {pages} {1440} (\bibinfo {year}
  {1977})}\BibitemShut {NoStop}%
\bibitem [{\citenamefont {Mohapatra}\ and\ \citenamefont
  {Senjanovic}(1978)}]{Mohapatra:1978fy}%
  \BibitemOpen
  \bibfield  {author} {\bibinfo {author} {\bibfnamefont {R.~N.}\ \bibnamefont
  {Mohapatra}}\ and\ \bibinfo {author} {\bibfnamefont {G.}~\bibnamefont
  {Senjanovic}},\ }\href {\doibase 10.1016/0370-2693(78)90243-5} {\bibfield
  {journal} {\bibinfo  {journal} {Phys.Lett.}\ }\textbf {\bibinfo {volume}
  {B79}},\ \bibinfo {pages} {283} (\bibinfo {year} {1978})}\BibitemShut
  {NoStop}%
\bibitem [{\citenamefont {Kuchimanchi}(1996)}]{Kuchimanchi:1995rp}%
  \BibitemOpen
  \bibfield  {author} {\bibinfo {author} {\bibfnamefont {R.}~\bibnamefont
  {Kuchimanchi}},\ }\href {\doibase 10.1103/PhysRevLett.76.3486} {\bibfield
  {journal} {\bibinfo  {journal} {Phys. Rev. Lett.}\ }\textbf {\bibinfo
  {volume} {76}},\ \bibinfo {pages} {3486} (\bibinfo {year} {1996})},\ \Eprint
  {http://arxiv.org/abs/hep-ph/9511376} {arXiv:hep-ph/9511376} \BibitemShut
  {NoStop}%
\bibitem [{\citenamefont {Mohapatra}\ and\ \citenamefont
  {Rasin}(1996)}]{Mohapatra:1995xd}%
  \BibitemOpen
  \bibfield  {author} {\bibinfo {author} {\bibfnamefont {R.~N.}\ \bibnamefont
  {Mohapatra}}\ and\ \bibinfo {author} {\bibfnamefont {A.}~\bibnamefont
  {Rasin}},\ }\href {\doibase 10.1103/PhysRevLett.76.3490} {\bibfield
  {journal} {\bibinfo  {journal} {Phys.Rev.Lett.}\ }\textbf {\bibinfo {volume}
  {76}},\ \bibinfo {pages} {3490} (\bibinfo {year} {1996})},\ \Eprint
  {http://arxiv.org/abs/hep-ph/9511391} {arXiv:hep-ph/9511391 [hep-ph]}
  \BibitemShut {NoStop}%
\bibitem [{\citenamefont {Babu}\ \emph {et~al.}(2002)\citenamefont {Babu},
  \citenamefont {Dutta},\ and\ \citenamefont {Mohapatra}}]{Babu:2001se}%
  \BibitemOpen
  \bibfield  {author} {\bibinfo {author} {\bibfnamefont {K.}~\bibnamefont
  {Babu}}, \bibinfo {author} {\bibfnamefont {B.}~\bibnamefont {Dutta}}, \ and\
  \bibinfo {author} {\bibfnamefont {R.}~\bibnamefont {Mohapatra}},\ }\href
  {\doibase 10.1103/PhysRevD.65.016005} {\bibfield  {journal} {\bibinfo
  {journal} {Phys.Rev.}\ }\textbf {\bibinfo {volume} {D65}},\ \bibinfo {pages}
  {016005} (\bibinfo {year} {2002})},\ \Eprint
  {http://arxiv.org/abs/hep-ph/0107100} {arXiv:hep-ph/0107100 [hep-ph]}
  \BibitemShut {NoStop}%
\bibitem [{\citenamefont {Mohapatra}\ \emph {et~al.}(1997)\citenamefont
  {Mohapatra}, \citenamefont {Rasin},\ and\ \citenamefont
  {Senjanovic}}]{Mohapatra:1997su}%
  \BibitemOpen
  \bibfield  {author} {\bibinfo {author} {\bibfnamefont {R.~N.}\ \bibnamefont
  {Mohapatra}}, \bibinfo {author} {\bibfnamefont {A.}~\bibnamefont {Rasin}}, \
  and\ \bibinfo {author} {\bibfnamefont {G.}~\bibnamefont {Senjanovic}},\
  }\href {\doibase 10.1103/PhysRevLett.79.4744} {\bibfield  {journal} {\bibinfo
   {journal} {Phys.Rev.Lett.}\ }\textbf {\bibinfo {volume} {79}},\ \bibinfo
  {pages} {4744} (\bibinfo {year} {1997})},\ \Eprint
  {http://arxiv.org/abs/hep-ph/9707281} {arXiv:hep-ph/9707281 [hep-ph]}
  \BibitemShut {NoStop}%
\bibitem [{\citenamefont {Kuchimanchi}(2012)}]{Kuchimanchi:2012xb}%
  \BibitemOpen
  \bibfield  {author} {\bibinfo {author} {\bibfnamefont {R.}~\bibnamefont
  {Kuchimanchi}},\ }\href {\doibase 10.1103/PhysRevD.86.036002} {\bibfield
  {journal} {\bibinfo  {journal} {Phys.Rev.}\ }\textbf {\bibinfo {volume}
  {D86}},\ \bibinfo {pages} {036002} (\bibinfo {year} {2012})},\ \Eprint
  {http://arxiv.org/abs/1203.2772} {arXiv:1203.2772 [hep-ph]} \BibitemShut
  {NoStop}%
\bibitem [{\citenamefont {Kuchimanchi}(2014)}]{Kuchimanchi:2012te}%
  \BibitemOpen
  \bibfield  {author} {\bibinfo {author} {\bibfnamefont {R.}~\bibnamefont
  {Kuchimanchi}},\ }\href {\doibase 10.1140/epjc/s10052-014-2726-5} {\bibfield
  {journal} {\bibinfo  {journal} {Eur. Phys. J. C}\ }\textbf {\bibinfo {volume}
  {74}},\ \bibinfo {eid} {2726} (\bibinfo {year} {2014})}\BibitemShut {NoStop}%
\end{thebibliography}%

\end{document}